\documentclass[pra,twocolumn,showpacs]{revtex4}
\usepackage{graphicx}

\begin{document}

\title{One-Way Entangled-Photon Autocompensating Quantum
Cryptography}

\author{Zachary~D.~Walton}
\email{walton@bu.edu} \homepage[Quantum Imaging Laboratory
homepage:~]{http://www.bu.edu/qil}

\author{Ayman~F.~Abouraddy}

\author{Alexander~V.~Sergienko}

\author{Bahaa~E.~A.~Saleh}

\author{Malvin~C.~Teich}
\affiliation{Quantum Imaging Laboratory, Departments of Electrical
\& Computer Engineering and Physics, Boston University, 8 Saint
Mary's Street, Boston, Massachusetts 02215-2421}


\begin{abstract}
A new quantum cryptography implementation is presented that uses
entanglement to combine one-way operation with an autocompensating
feature that has hitherto only been available in implementations
that require the signal to make a round trip between the users.
Using the concept of advanced waves, it is shown that this new
implementation is related to the round-trip implementation in the
same way that Ekert's two-particle scheme is related to the
original one-particle scheme of Bennett and Brassard.  The
practical advantages and disadvantages of the proposed
implementation are discussed in the context of existing schemes.
\end{abstract}

\pacs{03.65.Ud, 03.67.Dd, 42.50.Dv, 42.65.Ky}

\maketitle


The idea of using quantum systems for secure communications
originated in the 1970s with Stephen Wiesner's intuition that the
uncertainty principle, commonly derided as a source of noise,
could be harnessed to detect unauthorized monitoring of a
communication channel~\cite{Wiesner83}.  The first quantum
cryptographic protocol (BB84) was published by Charles H.~Bennett
and Giles Brassard in 1984~\cite{Bennett84}.  While rigorous
proofs of the security of BB84 under realistic conditions have
only recently emerged (cf. Ref.~\cite{Gottesman02} and references
therein), the ``no-cloning theorem''~\cite{Wootters82} published
in 1982 provides a one-line security proof applicable in ideal
circumstances. Given the obvious choice of light as a signal
carrier, the path to practical quantum cryptography was clear:
develop robust experimental methods to create, manipulate,
transmit, and detect single photons.  For an excellent summary of
progress in the theory and practice of quantum cryptography, see
Ref.~\cite{Gisin02}.

The nascent field of quantum cryptography took an unexpected turn
in 1992 when Artur Ekert published a new protocol~\cite{Ekert91}
that derived its security not from the impossibility of cloning a
quantum state, but rather the seemingly distinct phenomenon of the
violation of Bell's inequality~\cite{Bell64}.  The practical
importance of this scheme was immediately questioned by Bennett et
al.~\cite{Bennett92b}. They pointed out that the same hardware
required for Ekert's protocol could be used to implement the more
efficient BB84 protocol.  This is accomplished by regarding the
two-particle source together with one detection apparatus as a
single entity that produces a localized quantum state (i.e., one
of the four BB84 polarization states) to be detected by the other
detection apparatus. Although not described as such, their
objection amounted to an application of the concept of advanced
waves~\cite{Belinsky92}. This method, pioneered by David Klyshko,
establishes a formal equivalence between two optical constructs:
1) the propagation of two entangled photons from a localized
source to a pair of remote detectors, and 2) the propagation of a
single photon from one detector backwards towards the source,
where it is reflected, and then forward to the other detector. The
advanced-wave method is a powerful tool for developing intuition
about two-photon interference experiments that demonstrate
entanglement in time~\cite{Klyshko92}, space~\cite{Pittman95},
and, trivially, polarization.  For a discussion of apparent
backward-in-time processes in the more general context of quantum
information theory, see Ref.~\cite{Cerf97}.

The strong interest in absolutely secure communications has fueled
an ongoing effort to determine which protocol leads to the best
performance in practical implementations.  In 1997, Muller et
al.~introduced autocompensating quantum cryptography (AQC), in
which the optical signal makes a round trip between the legitimate
users (commonly referred to as Alice and Bob)~\cite{Muller97}. The
scheme is described as autocompensating since it provides
high-visibility interference without an initial calibration step
or active compensation of drift in the optical apparatus; these
favorable properties led the authors to refer to their scheme
informally as ``plug-and-play quantum cryptography.'' While this
scheme and its
refinements~\cite{Ribordy98,Bourennane00,Bethune02,Nishioka02,Stucki02}
represent substantial progress in the quest for a practical
quantum cryptography implementation, the requirement that the
signal travel both directions along the transmission line leads to
non-trivial technical difficulties.

In this article, we describe one-way entangled-photon
autocompensating quantum cryptography (OW-AQC) in which two
photons travel one way (e.g., from Alice to Bob), instead of one
photon traveling back and forth, as in AQC. The formal association
of OW-AQC with AQC follows directly from the advanced-wave view,
just as Ekert's scheme follows from BB84.  While Ekert's scheme
employs entanglement to allow an alternative space-time
configuration (signal source between Alice and Bob versus a source
on Alice's side), OW-AQC employs entanglement to achieve immunity
to interferometer drift within the original paradigm of a one-way
quantum channel from Alice to Bob.  Thus, our result provides a
new example of a capability afforded by quantum entanglement.

This article is organized as follows.  First, we briefly review
the standard AQC scheme.  Second, we introduce OW-AQC and show
that it combines one-way operation with the insensitivity to drift
that is characteristic of its predecessor.  Third, we point out
the formal equivalence of the two methods from the advanced-wave
viewpoint using space-time diagrams. Finally, we discuss the
relative merits of OW-AQC.


\begin{figure}
\includegraphics{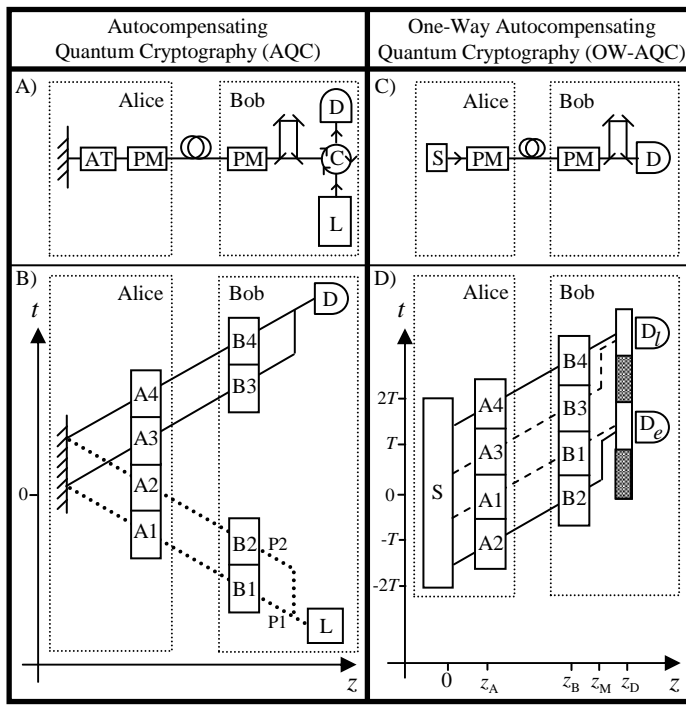}
\caption{A) and C) depict schematics for AQC and OW-AQC,
respectively. L is a source of laser pulses, S emits the
two-photon entangled state $|\Psi\rangle$ described by
Eq.~(\protect{\ref{qs}}), C is a circulator, AT is an attenuator,
PM is a phase modulator, and (D, $\mbox{D}_e$, $\mbox{D}_l$) are
detectors.  B) and D) depict the associated space-time diagrams
which indicate how the interference condition between the two
amplitudes is controlled by both Alice and Bob. The dotted
space-time traces in B) are used in the text to explain the
relationship between the two methods from the viewpoint of
advanced waves.  In D), the four rectangles at the point $z=z_D$
correspond to the four time intervals labeled at $z=z_A$ and
$z=z_B$.  The unshaded boxes indicate the two time intervals
during which Bob's detector is activated.  The solid and dashed
space-time traces depict two interfering two-photon amplitudes, as
described in the text.}\label{fig-1}
\end{figure}

Figure~\ref{fig-1}A contains a schematic of AQC. The protocol
begins with Bob launching a strong pulse from a laser (L) into a
Mach--Zehnder interferometer via a circulator (C). This
interferometer splits the pulse into an advanced amplitude (P1)
and a retarded amplitude (P2). The amplitudes travel through phase
modulators (PM) on Bob's side and Alice's side, and are then
attenuated (AT) to the single photon level and reflected by Alice
back to Bob. Although both P1 and P2 will again be split at Bob's
Mach--Zehnder interferometer, by gating his detector
appropriately, Bob can postselect those cases in which P1 takes
the long path and P2 takes the short path on the return trip.
Thus, the interfering amplitudes experience identical delays on
their round trip, ensuring insensitivity to drift in Bob's
interferometer.

The role of the phase modulators can be readily understood by
examining the space-time diagram of this protocol (see
Fig.~\ref{fig-1}B).  The eight boxes (A1--A4, B1--B4) refer to the
phase settings on the two modulators as the two amplitudes pass
through each of them twice.  For example, B2 refers to the phase
acquired by the delayed amplitude of the pulse that Bob sends to
Alice, while B4 refers to the phase acquired by the same amplitude
as it travels back from Alice to Bob.  It should be understood
that B1--B4 refer to settings of the same physical phase shifter
at different times (and similarly for A1--A4). The probability of
a detection at Bob's detector is given by
\begin{eqnarray}
P_{\mbox{d}}\propto
1&+&\cos[(\mbox{B2}-\mbox{B1})+(\mbox{A2}-\mbox{A1})\nonumber\\
&+&(\mbox{A4}-\mbox{A3})+(\mbox{B4}-\mbox{B3})].
\end{eqnarray}
From this expression we see that only the relative phase between
the phase modulator settings affects the probability of detection.
Thus, by setting $\mbox{B1}=\mbox{B2}$ and $\mbox{A1}=\mbox{A2}$,
Alice and Bob can implement the interferometric version of BB84 by
encoding their cryptographic key in the difference settings
$\Delta\phi_A\equiv\mbox{A4}-\mbox{A3}$ and
$\Delta\phi_B\equiv\mbox{B4}-\mbox{B3}$.  Since the resulting
expression
\begin{equation}\label{pnp}
P_{\mbox{d}}\propto1+\cos\left(\Delta\phi_A + \Delta\phi_B\right)
\end{equation}
is independent of the time delay in Bob's interferometer and the
absolute phase settings in either modulator,  Alice and Bob are
able to achieve high-visibility interference without initial
calibration or active compensation of drift.


Figure~\ref{fig-1}C contains a schematic of OW-AQC. Alice's source
(S) produces a specific two-photon state which is transmitted to
Bob and analyzed with a Mach--Zehnder interferometer and a single
detector that is activated for two distinct time intervals.  As in
AQC, Alice and Bob change the settings of their respective phase
modulators at specific time intervals in order to implement BB84.
The two-photon state that Alice sends to Bob consists of an early
photon (which is emitted from Alice's source in the time interval
$t_e\in[-2T,0]$) and a late photon (which is emitted in the time
interval $t_l\in[0,2T]$). The joint emission times of the early
photon and the late photon are described by the state
$|\Psi\rangle = \int \int dt_e dt_l\,\,
f(t_e,t_l)\,|t_e\rangle\,|t_l\rangle$, where
\begin{equation}\label{qs}
f(t_e,t_l)\propto\left\{\begin{array}{lll}\delta(t_e+t_l)&\quad&-2T< t_e<0\\
0 &\quad& \mbox{otherwise.}
\end{array}\right.
\end{equation}
This particular entangled state entails perfect anti-correlation
in the time of emission of the two photons; thus, while the
difference in emission time of the two photons is uniformly
distributed over the interval $[0,4T]$, the sum of the emission
times is fixed at $t=0$ for each emitted pair. By Fourier duality,
the two photons are correlated in frequency. While the typical
configurations for practical sources of entangled photon pairs
produce frequency anti-correlation, the frequency-correlated case
has been discussed in several
papers~\cite{Campos90,Keller97,Erdmann00,Kim02,Giovannetti02,Walton02c}.

Figure~\ref{fig-1}D presents a space-time diagram of the OW-AQC
protocol. The two-photon entangled state is sent through Alice's
phase modulator at position $z=z_A$ where she sets the phase
shifts (A1--A4) for the four time intervals indicated in the
diagram. Next, the two-photon state is transmitted along the
channel to Bob, where it is sent through Bob's phase modulator
($z=z_B$).  A Mach--Zehnder interferometer ($z=z_M$) then delays a
portion of the radiation by a time $\tau$. Finally, Bob's detector
($z=z_D$) is activated for two time intervals of length $T$ that
correspond to the second halves of the early and late photon wave
packets.  Gating Bob's detector in this way postselects the cases
in which the advanced (delayed) portion of each photon takes the
long (short) path.  This postselection reduces the photon flux by
half and obviates the need for rapid switching of optical paths.
Since the time intervals are non-overlapping, we may consider that
Bob is using two detectors that are distinguished by the ordering
of their respective time windows. Thus, for the rest of the
letter, we refer to two detectors on Bob's side, $\mbox{D}_e$ and
$\mbox{D}_l$, which correspond respectively to the early and late
activation intervals of Bob's single physical detector.

The two-photon interference can be seen by examining the
space-time trajectories of two specific two-photon amplitudes.  In
Fig.~\ref{fig-1}D, the solid space-time traces entail emission
times $(t_e,t_l)=(-3T/2,3T/2)$ and the dashed traces entail
emission times $(t_e,t_l)=(-T/2,T/2)$.  For delay $\tau=T$, the
portion of the solid and dashed amplitudes leading to a
coincidence are indistinguishable after Bob's Mach--Zehnder
interferometer. This indistinguishability brings about quantum
interference that varies continuously between completely
constructive and completely destructive, depending on the joint
phase settings A1--A4, B1--B4.

By activating detectors $\mbox{D}_e$ and $\mbox{D}_l$ for a
duration $T$ at times $\frac{z_D}{c}-T$ and $\frac{z_D}{c}+T$,
respectively, Bob establishes the following relation between the
electric-field operators $\hat{E}_{e,l}$ at his detectors and the
annihilation operator $\hat{a}(t)$ associated with $|t\rangle$,
\begin{widetext}
\begin{eqnarray}
\hat{E}_e(t_1)&\propto&\left\{\begin{array}{lll}e^{i\left(\mbox{A2}+\mbox{B2}\right)}\hat{a}(t_1-\frac{z_D}{c}-\tau)+e^{i\left(\mbox{A1}+\mbox{B1}\right)}\hat{a}(t_1-\frac{z_D}{c})&\quad& -T<t_1-\frac{z_D}{c}<0\\ 0 &\quad&\mbox{otherwise} \\ \end{array}\right.\\
\hat{E}_l(t_2)&\propto&\left\{\begin{array}{lll}e^{i\left(\mbox{A3}+\mbox{B3}\right)}\hat{a}(t_2-\frac{z_D}{c}-\tau)+e^{i\left(\mbox{A4}+\mbox{B4}\right)}\hat{a}(t_2-\frac{z_D}{c})&\quad&
T<t_2-\frac{z_D}{c}<2T\\ 0 &\quad& \mbox{otherwise,} \\
\end{array}\right.
\end{eqnarray}
\end{widetext}
where $\tau$ is the delay in Bob's Mach--Zehnder interferometer,
and $c$ is the speed of light.  Substituting into the expression
for the probability of a coincidence $ P_{\mbox{c}}\propto\int\int
dt_1 dt_2\, |\langle
0|\hat{E}_e(t_1)\hat{E}_l(t_2)|\Psi\rangle|^2$, we obtain
\begin{eqnarray}\label{owpnp}
P_{\mbox{c}}&\propto&\Lambda\left(\frac{\tau-T}{T}\right)\left[1+\cos\left(\Delta\phi_A
+ \Delta\phi_B\right)\right]\nonumber \\
& + &
\frac{1}{2}\left[\Lambda\left(\frac{\tau-T/2}{T/2}\right)+\Lambda\left(\frac{\tau-3T/2}{T/2}\right)\right]\quad,
\end{eqnarray}
where $\Lambda(x)=1-|x|$ for $-1<x<1$ and $0$ otherwise. When
$\tau=T$, this equation reduces to the expression for the
probability of detection in AQC [see Eq.~(\ref{pnp})].  To
implement the interferometric version of BB84, Alice and Bob hold
the settings of their respective phase modulators constant for the
first two time intervals depicted in Fig.~\ref{fig-1}D (i.e.,
$A1=A2$ and $B1=B2$), and manipulate the difference terms,
$\Delta\phi_A\equiv\mbox{A4}-\mbox{A3}$ and
$\Delta\phi_B\equiv\mbox{B4}-\mbox{B3}$.  The crucial point is
that the interference condition is independent of the absolute
setting or drift in either of the phase modulators. This
demonstrates that OW-AQC achieves the insensitivity to absolute
phase settings characteristic of AQC, while requiring only one
pass through the optical system.


It is instructive to compare the space-time diagrams in
Figs.~\ref{fig-1}B and~\ref{fig-1}D.  Reflecting the dotted traces
in Fig.~\ref{fig-1}B around the line $t=0$ results in the exact
space-time arrangement of Fig.~\ref{fig-1}D. This construction
also provides a clear explanation of why the two-photon state
described by Eq.~(\ref{qs}) is chosen to possess frequency
correlation instead of the more common frequency anti-correlation.
A device that creates pairs of photons with coincident frequencies
(S in Fig.~\ref{fig-1}D) is nothing more than a mirror (as
required by Fig.~\ref{fig-1}B) when analyzed from the
advanced-wave viewpoint. Thus, Klyshko's advanced-wave
interpretation provides an intuitive justification for the
equivalence between the probability of single-photon detection
[Eq.~(\ref{pnp})] and the probability of two-photon coincidence
[Eq.~(\ref{owpnp})] with respect to the phase modulator settings
A1--A4 and B1--B4.


Here we provide a qualitative comparison of AQC and OW-AQC.  While
AQC requires that only one photon travel the distance between
Alice and Bob after Alice attenuates Bob's signal to the
single-photon level, OW-AQC requires that two photons travel the
same distance. Thus, the loss incurred in OW-AQC is approximately
twice that of AQC for the same distance.  However, the use of a
strong pulse on the first leg of the round trip in AQC also
contributes to a disadvantage relative to OW-AQC. Specifically,
backscattered light from the strong pulse is guided directly into
Bob's detectors and can lead to unacceptably high bit-error rates.
Another advantage of OW-AQC is immunity from the ``Trojan horse
attack''~\cite{Gisin02}, in which Eve sends an optical signal into
Alice's lab and measures the state of the reflected light in order
to infer the setting of Alice's phase modulator. While an optical
isolator can subvert this attack in the case of OW-AQC, the
bidirectional flow of optical signals in AQC prevents this
defence.  In AQC, the probability of detection is independent of
the delay $\tau$ in Bob's interferometer [see Eq.~(\ref{pnp})],
while in OW-AQC, the interference condition is independent of
$\tau$, but the visibility of this interference is not [see
Eq.~(\ref{owpnp})]. Thus, while drift in the absolute values of
the phase modulations will not affect the performance of OW-AQC,
drift in the optical delay must be minimized to maintain
high-visibility interference.

It is important to note that OW-AQC requires the
frequency-correlated two-photon entangled state described in
Eq.~(\ref{qs}). This state has been investigated
theoretically~\cite{Campos90}, and several experimental methods
for creating the state have been
proposed~\cite{Keller97,Erdmann00,Kim02,Giovannetti02,Walton02c}.
However, the state has not yet been experimentally demonstrated.
While frequency-anticorrelated photon pairs are naturally
generated when a monochromatic pump beam impinges on a nonlinear
crystal, frequency-correlated photon pairs are only generated when
a broad-band pump is used, and constraints on the phase {\em and}
group velocities of the pump, signal, and idler are satisfied.
These constraints can be satisfied in a collinear setup by
exploiting the birefringence of the nonlinear
crystal~\cite{Keller97,Erdmann00}. Enhanced flexibility in
satisfying these constraints can be achieved by imposing a
periodic modulation of the crystal's nonlinear
coefficient~\cite{Giovannetti02}.  A second approach to satisfying
these constraints is to exploit the inherent symmetry of a
configuration in which a nonlinear waveguide is pumped at normal
incidence such that the down-converted photons are
counter-propagating~\cite{Walton02c}.  The advantage of this
method is that frequency-correlated photon pairs can be generated
regardless of the dispersion characteristics of the nonlinear
material.


In summary, we have described a new quantum cryptography
implementation that exploits quantum entanglement to achieve the
favorable stability of AQC without requiring a round trip between
Alice and Bob.  This work represents the first demonstration that
quantum entanglement can offer practical advantages with respect
to noise in quantum cryptography implementations.  The next step
in evaluating the promise of this approach for practical quantum
cryptography involves explicit experimental proposals for creating
the source described by Eq.~(\ref{qs}) and quantitative
performance analysis.

Both this work and Ekert's landmark paper~\cite{Ekert91} linking
quantum cryptography and Bell's theorem describe two-photon
interference effects that employ novel space-time configurations
to perform tasks previously achieved with single-photon
interference. These constructions can be seen as applications of
Klyshko's theory of advanced waves, which provides a formal
equivalence of one- and two-photon interference experiments.

\begin{acknowledgments}
This work was supported by the National Science Foundation; the
Center for Subsurface Sensing and Imaging Systems (CenSSIS), an
NSF Engineering Research Center; and the Defense Advanced Research
Projects Agency (DARPA).

\end{acknowledgments}

\end{document}